\documentclass[%
 reprint,
 amsmath,amssymb,
 aps,
]{revtex4-2}

\usepackage{graphicx}% Include figure files
\usepackage{dcolumn}% Align table columns on decimal point
\usepackage{bm}% bold math
\usepackage{lineno}
\usepackage{siunitx}
\usepackage{textcomp}
\begin{document}

\preprint{APS/123-QED}

\title{All-optical control of topological valley transport in graphene metasurfaces}

\author{Yupei Wang}
\email{yupei.wang.18@ucl.ac.uk}
\author{Jian Wei You}
\altaffiliation[Also at ]{State Key Laboratory of Millimeter Waves, Department of Information Science and Engineering, Southeast University, Nanjing, 211189, China.}
\author{Nicolae C. Panoiu}
\affiliation{Department of Electronic and Electrical Engineering, University College London, London WC1E 7JE, U.K. \\}

\date{\today}% It is always \today, today,

\begin{abstract}
We demonstrate that the influence of Kerr effect on valley-Hall topological transport in graphene metasurfaces can be used to implement an all-optical switch. In particular, by taking advantage of the large Kerr coefficient of graphene, the index of refraction of a topologically-protected graphene metasurface can be tuned via a pump beam, which results in an optically controllable frequency shift of the photonic bands of the metasurface. This spectral variation can in turn be readily employed to control and switch the propagation of an optical signal in certain waveguide modes of the graphene metasurface. Importantly, our theoretical and computational analysis reveals that the threshold pump power needed to optically switch ON/OFF the signal is strongly dependent on the group velocity of the pump mode, especially when the device is operated in the slow-light regime. This study could open up new routes towards active photonic nanodevices whose underlying functionality stems from their topological characteristics.
\end{abstract}

\maketitle

%\tableofcontents

\section{Introduction}
Topological plasmonic edge states have recently attracted a great deal of attention, particularly due to their unique and robust optical properties, including unidirectional propagation and backscattering-free optical interaction with structural disorder \cite{1,2topological,3topological}. Topologically protected edge modes can emerge inside a nontrivial bandgap, which is usually generated by gapping out a symmetry-protected Dirac cone. For example, time-reversal symmetry of photonic structures can be broken by applying an external static magnetic field, leading to an analogue quantum-Hall effect \cite{1,2topological,3topological,4topologicalano}. In addition to breaking the time-reversal symmetry, the spatial-inversion symmetry can also be broken via spatial perturbations that are not invariant to the inversion symmetry transformation, resulting in an analogue quantum valley-Hall effect \cite{1,2topological,3topological}. By emulating in photonics the valley degree of freedom introduced in solid state physics \cite{valley,5valleygraphene}, it has been suggested that two-dimensional (2D) honeycomb photonic lattices exhibit a valley-dependent topological index, the so-called valley Chern number. This topological invariant is expressed in terms of the integral of the Berry curvature over the vicinity of the valleys located at the $K,\,K^{\prime}$ symmetry points \cite{sivalley}. When the difference across an interface between the valley Chern numbers of two valley-Hall topological photonic crystals (PhCs) is different from zero, topologically-protected valley modes can emerge at and propagate along the domain-wall interface separating the two PhCs with different topology \cite{6valley}.

To date, most of the studies of topological photonics have focused on passive optical systems operating in the linear regime \cite{7nonlinearreview}. However, introducing optical nonlinear effects to topological systems can lead to active photonic devices with new or improved functionalities, including ultra-short pulsed lasers, optical signal processing, and ultra-fast all-optical switches \cite{8nonlinearwork,9ultrafast,10nonlinear}. To this end, applications of nonlinear topological photonics, such as topological lasers \cite{laser1,laser2}, lattice edge solitons \cite{soliton1,soliton2}, optically tunable mode couplers \cite{modecouple}, and frequency convertors \cite{SH,SHL,THG,FWM}, have already been proposed theoretically and in some cases successfully implemented in a variety of experimental platforms, including optical waveguides, optical resonators, and metamaterials \cite{Nonlineartopological}. Specifically, the combination of the Kerr nonlinearity and nontrivial topological characteristics can potentially lead to a new class of active photonic devices where the system topology plays a key r\^{o}le. Since the Kerr coefficient of graphene is several orders of magnitude larger than that of bulk optical materials commonly used in practice \cite{11graphene}, graphene-based topological platforms can have a major impact to the development of advanced ultra-fast active photonic devices.

In this article, we demonstrate that an optical signal propagating in a topologically-protected valley mode of a graphene plasmonic crystal waveguide can be optically controlled through the Kerr effect induced by a pump beam injected in a bulk mode of the plasmonic crystal. In particular, by introducing additional holes with properly chosen size in a graphene metasurface with hexagonal lattice, so as it is no longer invariant to spatial-inversion symmetry transformations, one can create a nontrivial bandgap that originates from a symmetry-protected Dirac cone. We show that, as a result, topological interface modes presenting the familiar property of unidirectional light propagation are generated inside the nontrivial frequency bandgap. Employing this unidirectionality feature in conjunction with the large nonlinear refractive index coefficient of graphene, it is demonstrated that an optical signal propagating in the proposed topological waveguide can be optically controlled via an optical pump beam propagating in one of the bulk modes of the metasurface. Moreover, our study indicates that the pump power required to switch ON/OFF the optical signal is strongly dependent on the group-velocity (GV) of the pump, being significantly reduced when the proposed optical switch operates in the slow-light (SL) regime.

The article is organized as follows. In the next section we present the geometrical configuration and properties of the photonic band structure of the proposed graphene plasmonic crystal. Then, in Section 3, we describe the operation of the optical switch and analyze quantitatively its optical response. The main conclusions of our study are presented in the last section of the article.

\section{Band analysis of graphene plasmonic waveguide}

\begin{figure}[b!]
\centering
\includegraphics[width=0.47\textwidth]{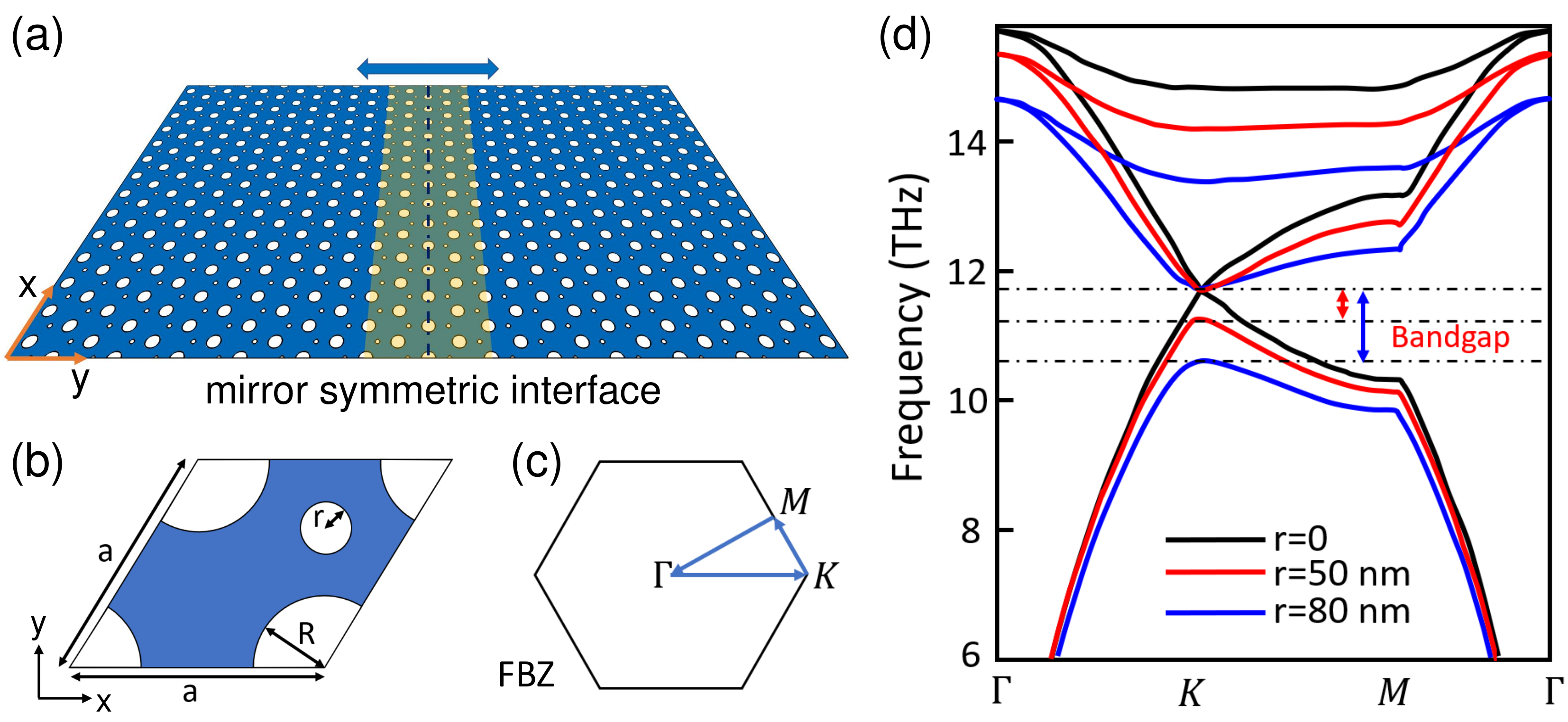}
\caption{(a) Schematic of the graphene plasmonic nanohole crystal with a domain-wall interface oriented along the $x$-axis. The left- and right-hand side domains are placed in a mirror-symmetric manner and contain the same unit cell. (b) Primitive unit cell of the graphene crystal with a lattice constant $a$, containing two air nanoholes with different radii $R$ and $r$. (c) First Brillouin zone of the graphene plasmonic crystal showing the high-symmetry points $\Gamma$, $K$, and $M$. (d) Band diagram of the graphene plasmonic crystal determined for different radii $r$.}
\label{cap1}
\end{figure}

The schematic of the proposed graphene metasurface is presented in Fig.~\ref{cap1}(a). It consists of a graphene sheet in which air nanoholes with two different radii are etched. The left and right semi-infinite metasurface domains are placed in a mirror-symmetric manner so as to generate a domain-wall interface along the $x$-axis, as illustrated by the yellow region in Fig.~\ref{cap1}(a). Each domain contains the same unit cell, whereby two air holes with different radii, $R$ and $r$, are arranged in a hexagonal array. The unit cell and the first Brillouin zone (FBZ) are shown in Figs.~\ref{cap1}(b) and \ref{cap1}(c), respectively. In this study, we fix the lattice constant, $a=400\sqrt{3}$~nm, and the radius of one of the nanoholes, $R=\SI{140}{\nm}$. The optical properties of graphene are generally described by its surface conductivity, given by Kubo's formula \cite{12Kubo,13permitivity}.

The band diagram of the graphene plasmonic crystal is calculated by the finite-element method (FEM) of Wave Optics Module in COMSOL\textsuperscript{\textregistered} Multiphysics 5.6 \cite{comsol} and validated using Synopsys\textsuperscript{\textregistered} BandSOLVE tool \cite{Synopsys}. As shown in Fig.~\ref{cap1}(d), the band diagrams were determined for different radii $r$. When $r=0$, as per the black curves in Fig.~\ref{cap1}(d), there is a Dirac cone at the $K$ point, which is protected by the $C_{6v}$ symmetry of the hexagonal arrangement of holes with radius $R$. The first and second bands indeed linearly cross at Dirac points located at the frequency of \SI{11.8}{\THz}. In order to open up the inversion-symmetry-protected Dirac cone, additional air holes ($r\neq0$) are introduced into the graphene plasmonic crystal. This reduces the spatial-inversion symmetry to the $C_{3v}$ point symmetry group \cite{14spatical-symmetry}. As a consequence, the $C_{6v}$-symmetry-protected Dirac cone is gapped out, resulting in the formation of a nontrivial frequency bandgap, as illustrated by the red and blue curves in Fig.~\ref{cap1}(d). Moreover, the variation of the width of the bandgap indicates its dependence on the radius of the added holes, $r$. Specifically, when $r$ increases from $\SIrange{50}{80}{\nm}$, the spectral width of the bandgap increases by about $\SI{1}{\THz}$. This frequency shift is almost entirely due to a blue-shift of the first band. In our subsequent computations, we consider graphene metasurfaces with radius $r$ fixed to $r=\SI{50}{\nm}$. This choice is guided by the fact that in this way one can achieve a relatively large nontrivial bandgap, which extends from $\SIrange{11.1}{11.8}{\THz}$.

\begin{figure}[t!]
\centering
\includegraphics[width=0.47\textwidth]{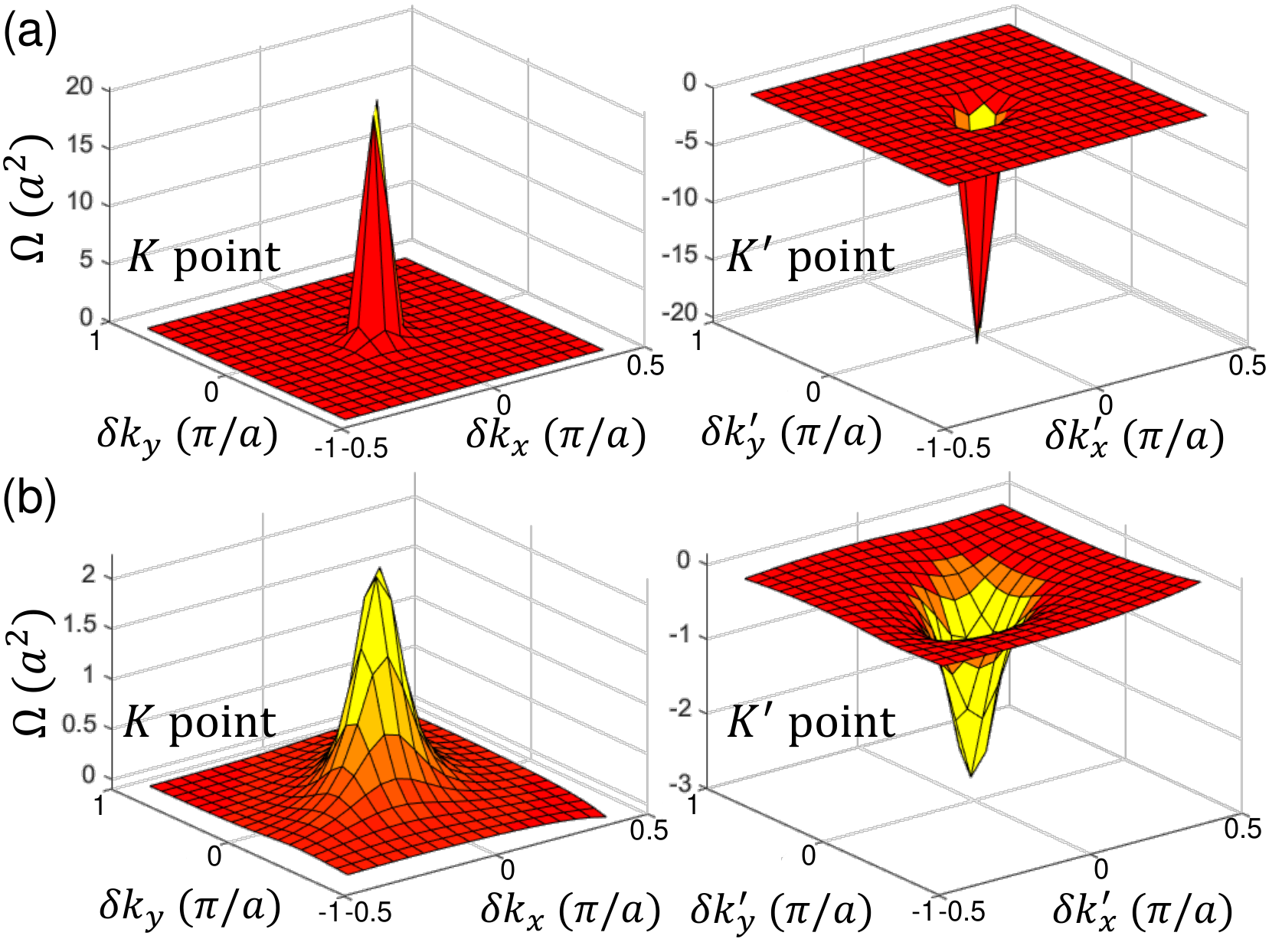}
\caption{Computed Berry curvature distribution of the first band over the $K$ and $K^{\prime}$ valleys, determined for the radius of the additional hole (a) $r=\SI{20}{\nm}$ and (b) $r=\SI{50}{\nm}$.}
\label{chernfigure}
\end{figure}

Since the lattice of holes has hexagonal symmetry, it possesses two nonequivalent valleys in the reciprocal space. The integral of the Berry curvature around each of these valleys defines the valley Chern number, which among other things determines the number of possible topological edge states \cite{6valley}. In the wave vector space, around each valley, the distribution of the Berry curvature around the two nonequivalent high-symmetry points $K$ and $K^{\prime}$ defines two valley-dependent Chern numbers, which can be computationally evaluated by the Wilson-loop approach in a discretized Brillouin zone \cite{Wilsonloopm}. To this end, valley Chern numbers of the first band [see Fig.~\ref{cap1}(d)] over the $K$ and $K^{\prime}$ valleys have been computed based on the FEM implemented in COMSOL. The results of these calculations, determined over a $18\times18$ computational grid covering the $k$-space, are plotted in Fig.~\ref{chernfigure}.

We show in Fig.~\ref{chernfigure}(a) the Berry curvature distribution of the first band when $r=\SI{20}{\nm}$ over the $K$ and $K^{\prime}$ valleys. As expected, the distribution of Berry curvature shows a sharp peak with opposite signs at $K$ and $K^{\prime}$ points. Specifically, the computed valley Chern numbers of the first band when $r=\SI{20}{\nano\meter}$ over $K$ and $K^{\prime}$ valleys are $0.49$ and $-0.49$, respectively. The small deviation of $0.01$ from the theoretical value of valley Chern number, $|C|=1/2$, illustrates the accuracy of the Wilson-loop computational method \cite{chern}. For $r=\SI{50}{\nm}$ corresponding to a wider nontrivial bandgap in Fig.~\ref{cap1}(d), the computed valley Chern numbers of the first band over $K$ and $K^{\prime}$ valleys decrease to $0.34$ and $-0.34$, respectively, as indicated in Fig.~\ref{chernfigure}(b). Note that when the frequency bandgap is relatively large, the distribution of the Berry curvature over the corners at $K$ and $K^{\prime}$ symmetry points of the Brillouin zone spreads significantly and overlaps with the Berry curvature distributions at neighbouring $K$ and $K^{\prime}$ symmetry points. This makes it more difficult to properly define the valley Chern number, as its calculation cannot be extended to the entire FBZ. Therefore, the theoretical valley Chern number $C=\pm1/2$ can only be obtained close to the Dirac point with an infinitesimally small perturbation \cite{wong2020gapless}.

Following this analysis, and for the sake of simplicity of our further discussions, we fix the valley Chern number of the first band over the $K$ and $K^{\prime}$ valleys to $C_K=\pm1/2$ and $C_{K^{\prime}}=\mp1/2$, respectively, values that correspond to an infinitesimally small perturbation. As a result, when a domain-wall interface with valley Chern number difference $\Delta C=\pm1$ is created by rotations of the graphene crystal by $\pi/3$, $\pi$, or $5\pi/3$, a pair of valley modes is generated inside the bandgap. These topological modes propagate along the domain-wall interface in opposite direction, namely they have positive and negative GV.
\begin{figure}[t!]
\centering
\includegraphics[width=0.47\textwidth]{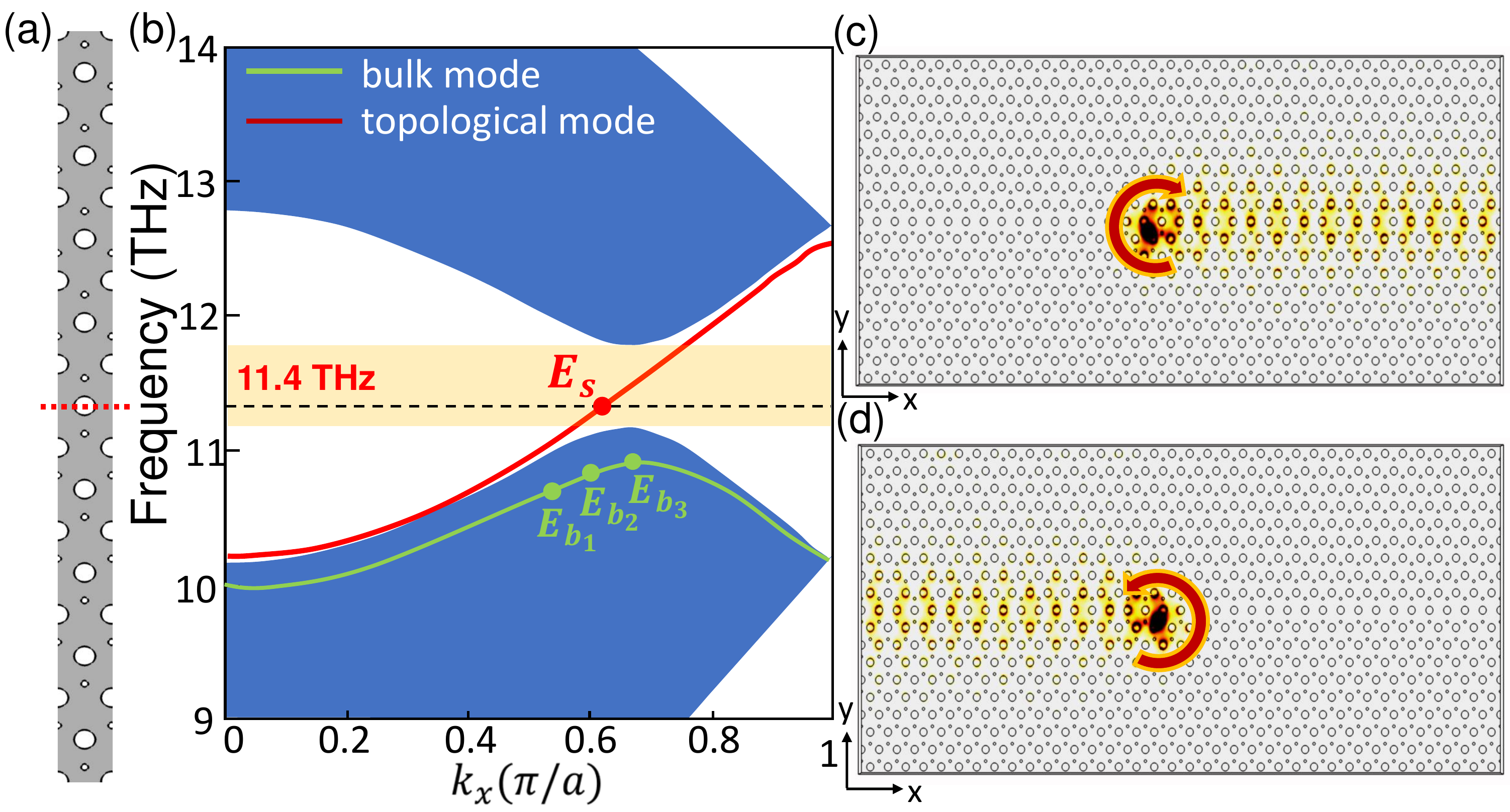}
\caption{(a) Supercell of a finite graphene plasmonic waveguide with a domain-wall interface constructed from two mirror-symmetric domains. (b) Projected band diagram of the graphene plasmonic crystal waveguide corresponding to $r=\SI{50}{\nm}$. A topologically protected interface-mode band (red line) exists inside the nontrivial bandgap. (c), (d) Unidirectional propagation of topological interfacial modes along the positive (negative) direction of the $x$-axis, when the waveguide is excited by a LCP (RCP) source at \SI{11.4}{\THz}, respectively.}
\label{cap2}
\end{figure}

We now consider two semi-infinite graphene metasurfaces placed together after one of them is rotated by $\pi$, thus forming a mirror-symmetric domain-wall interface, as shown in Fig.~\ref{cap2}(a). The finite valley-dependent Chern number makes it possible to form valley-momentum locked interfacial modes, illustrated in the projected band diagram of Fig.~\ref{cap2}(b). The blue regions represent the bulk modes, whereas a topological interface-mode band inside the nontrivial bandgap is marked by a red curve. In order to gain deeper physical insights into the properties of this topological mode, the corresponding field distribution is calculated at the frequency of \SI{11.4}{\THz}, which corresponds to the middle of the bandgap. The particular topological mode is indicated by the symbol $E_s$. As illustrated in Fig.~\ref{cap2}(c), this topological mode propagates along the domain-wall interface and is highly confined at the interface.

Furthermore, the unidirectional feature of valley-momentum locked interface modes is also investigated. To this end, six electric dipoles with increasing or decreasing phase difference ($\pm \pi/3$) are placed together at the corners of a small hexagon, thus generating light with specific chirality. To be more specific, under the excitation of a left circularly-polarized (LCP) source at the frequency of \SI{11.4}{\THz}, see Fig.~\ref{cap2}(c), the topologically protected interfacial mode propagates along the domain-wall interface in the positive direction of the $x$-axis with a positive group velocity, whereas the opposite is true for the topological interfacial mode excited by a right circularly-polarized (RCP) source with the same frequency, as per Fig.~\ref{cap2}(d).

\section{Active all-optically controlable switch}
Taking advantage of the unidirectional feature of the topological interface mode of the proposed graphene waveguide as well as the strong optical nonlinearity of graphene, a graphene-based all-optical switch is designed. To be more specific, we study the influence of the Kerr effect in graphene on the propagation characteristics of topological interface modes. As a versatile 2D material with large relaxation time (low optical losses) at THz frequencies, optical near-field enhancement \cite{15graphene}, and large optical nonlinearity \cite{16graphenenonlinear}, graphene is ideally suited for nonlinear optics applications.

The refractive index variation induced by the Kerr effect in the graphene metasurface, in response to the electric field generated by an optical pump, is given by \cite{10nonlinear}:
\begin{equation}\label{eq:1}
\Delta n_g(\mathbf{r})=\frac{1}{2}c\epsilon_0n_gn_2\vert \mathbf{E}_p(\mathbf{r})\vert^2,
\end{equation}
where $n_g$ is the refractive index of graphene, $n_2=\SI{7.5e-11}{\meter^2/\watt}$ is the nonlinear refractive coefficient of graphene \cite{17n2ofgraphene}, and $\mathbf{E}_p(\mathbf{r})$ is the field amplitude of the pump. Moreover, the pump power per unit cell, $P_p$, carried by a bulk mode $\mathbf{E}_b$ is given by \cite{18modepower}:
\begin{equation}\label{eq:2}
P_p=\frac{v_g}{4a}\int_{V_{cell}}\frac{\partial}{\partial\omega}[\omega\epsilon_{\omega}(\mathbf{r})]\vert \mathbf{E}_b(\mathbf{r})\vert^2dV,
\end{equation}
where $v_g=d\omega/dk_x$ is the GV, $V_{cell}$ is the volume of the unit cell, and $\epsilon_{\omega}(\mathbf{r})$ is the frequency-dependent permittivity distribution.
\begin{figure}[b!]
\centering
\includegraphics[width=0.47\textwidth]{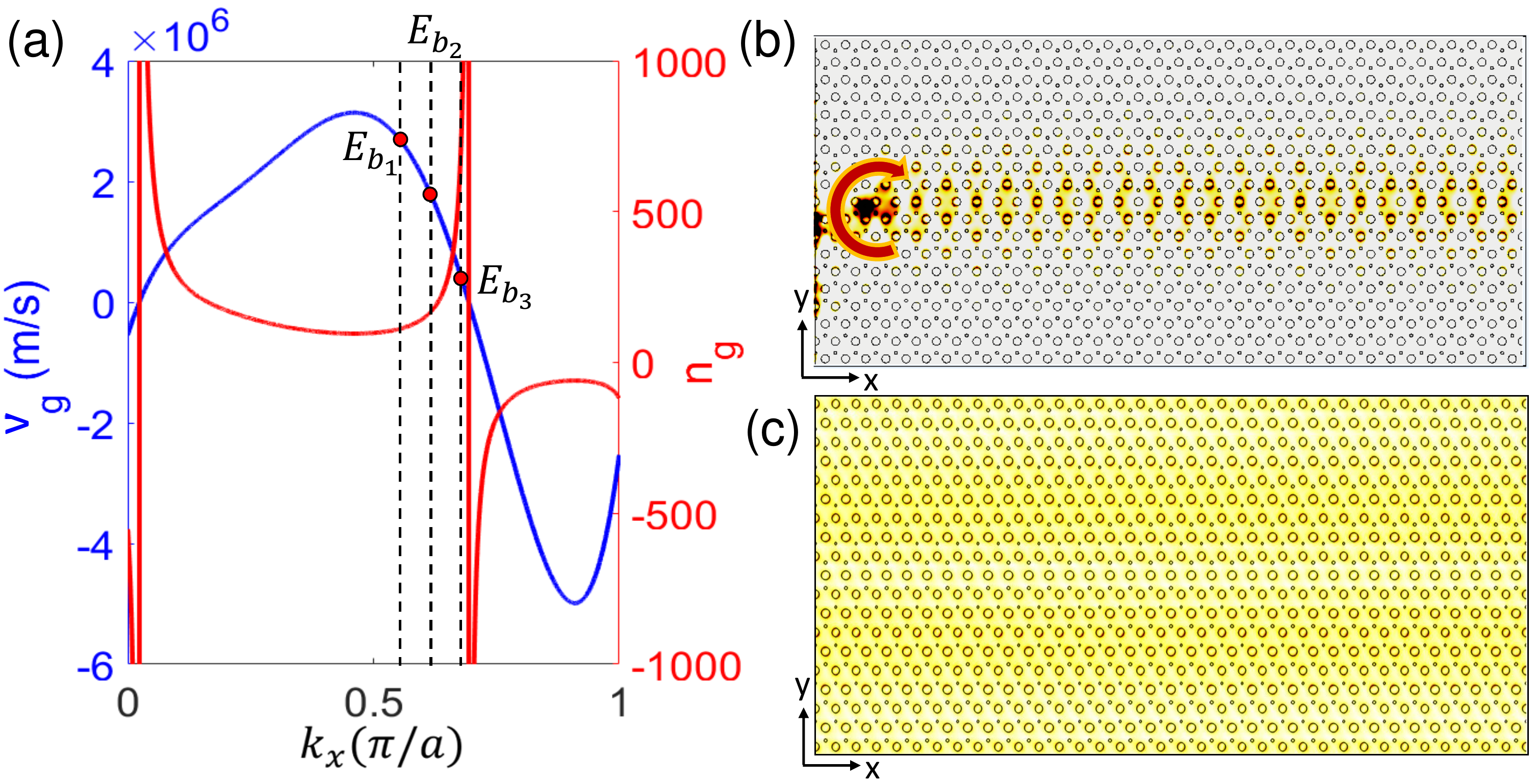}
\caption{(a) Dependence of group velocity $v_g$ and group index $n_g=c/v_g$ on $k_x$, corresponding to the bulk mode containing $E_{b_1}$, $E_{b_2}$, and $E_{b_3}$ [(green line in Fig.~\ref{cap2}(b)]. (b) Field profile of the topological interface mode at $\SI{11.4}{\tera\hertz}$, indicated by $E_{s}$ in Fig.~\ref{cap2}(b). (c) Field distribution of the pump injected in the bulk mode $E_{b_1}$.}
\label{cap3}
\end{figure}

To investigate the influence of the pump induced Kerr effect on the light propagation in the topological mode, several bulk modes with different GV are employed as pump modes. Since the power of the signal is much smaller than that of the pump, the contribution of the signal to the Kerr induced variation of the refractive index is neglected. As a result, the signal propagation in the topological mode is solely controlled by the optical pump.

The optical characteristics of the pump and signal modes are summarized in Fig.~\ref{cap3}. For the bulk mode band indicated by the green curve in Fig.~\ref{cap2}(b), the GV $v_g$ and group index, $n_g=c/v_g$, are shown in Fig.~\ref{cap3}(a). The three bulk modes, $E_{b_{1}}$ [$k_x=0.55(\pi/a)$], $E_{b_{2}}$ [$k_x=0.61(\pi/a)$], and $E_{b_{3}}$ [$k_x=0.67(\pi/a)$], are chosen in such a way that the first two are fast-light (FL) modes, whereas the last one is located in the SL region (the corresponding group index approaches $1000$). Slow light can generally provide strong field enhancement and consequently can lead to increased nonlinear optical interactions \cite{19slowlight}. Moreover, the corresponding field distributions of the signal and pump modes are plotted in Figs.~\ref{cap3}(b) and \ref{cap3}(c), respectively. The field distribution of the topological valley mode $E_s$, depicted in Fig.~\ref{cap2}(b), shows that a signal generated by a LCP source with \SI{11.4}{\THz} is tightly confined along the domain-wall interface. Moreover, the spatial field profile of the bulk mode $E_{b_{1}}$ with the largest GV among the three chosen bulk modes reveals a relatively uniform field distribution that extends across the entire graphene metasurface. Note that our simulations produced similar electric field distributions for the other two bulk modes, $E_{b_{2}}$ and $E_{b_{3}}$, the main difference being the value of the field amplitude corresponding to a given pump power.
\begin{figure}[t!]
\centering
\includegraphics[width=0.47\textwidth]{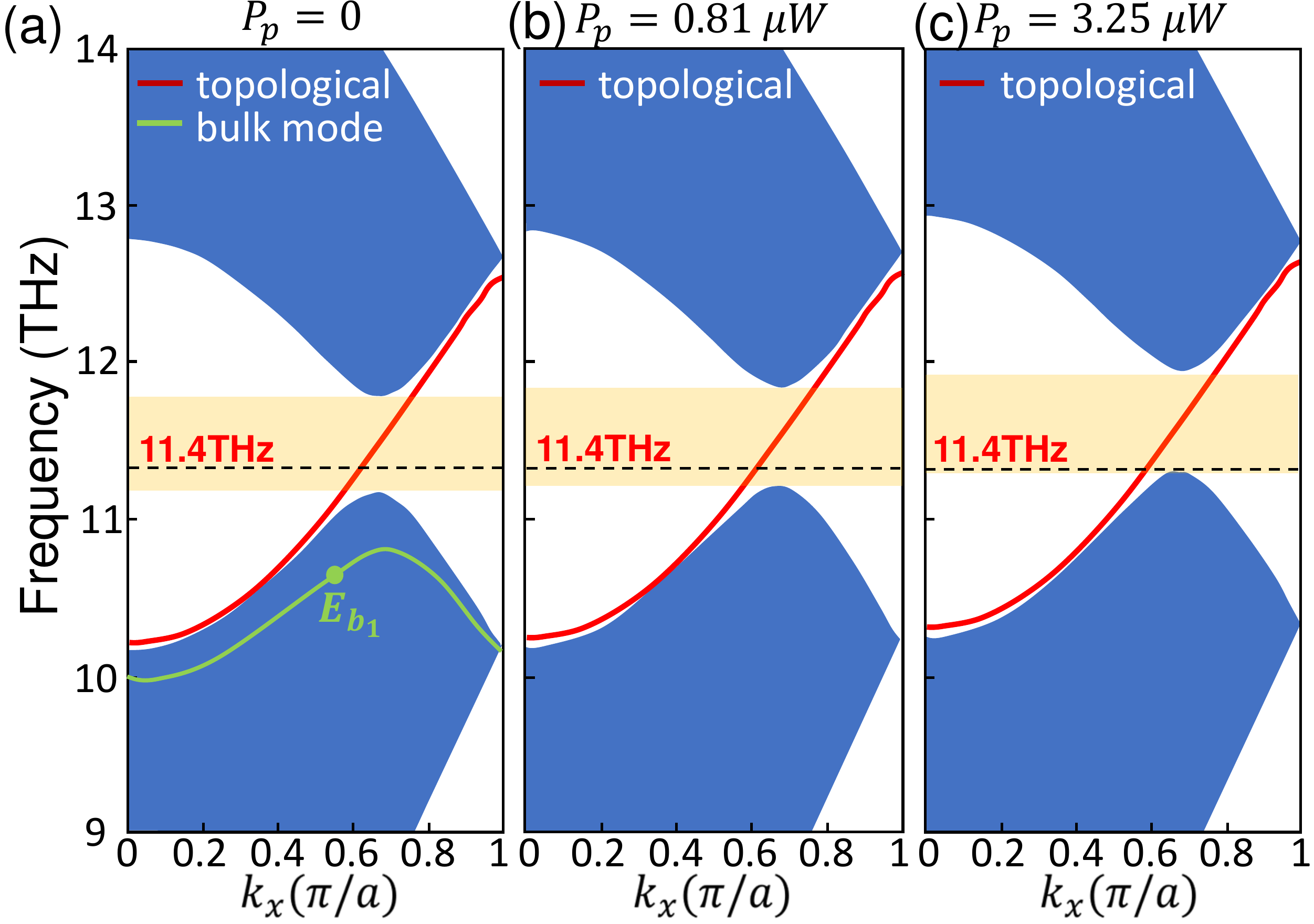}
\caption{(a), (b), (c) Projected band diagram of graphene nanohole plasmonic waveguide corresponding to the pump injected in the bulk mode $E_{b_{1}}$, determined for the pump power $P_p=0$, $P_p=\SI{0.81}{\uW}$, and $P_p=\SI{3.25}{\uW}$, respectively.}
\label{cap4}
\end{figure}

Since the optical Kerr effect in the graphene metasurface is induced by the pump, we investigated the dependence of the projected photonic band diagram on the pump power, $P_p$, the corresponding results being displayed in Fig.~\ref{cap4}. In order to determine the projected band structure corresponding to a pump power, $P_p$, we proceeded as follows. First, we computed the field distribution of the pump mode, $\mathbf{E}_b(\mathbf{r})$, and its GV, $v_{g}$, then scaled it using Eq.~\eqref{eq:2} in such a way that the optical power carried by the mode was $P_p$. Subsequently, using Eq.~\eqref{eq:1}, we calculated the Kerr-induced variation of the index of refraction of graphene and the corresponding distribution of the electric permittivity of the graphene metasurface. In the final step, this nonlinear permittivity was employed to determine the nonlinearly modified projected photonic band diagram of the metasurface. It should be noted that since the pump mode is periodic along the $x$-axis, a projected band structure can be defined in the optically pumped metasurface, too.

In this analysis, we assumed that the pump power $P_p$ was equal to 0 (linear regime), \SI{0.81}{\uW}, and \SI{3.25}{\uW}. The results of these computations indicate that as the pump power increases the frequency of the topological bandgap is blue-shifted by as much as \SI{0.3}{\THz}. Therefore, if the frequency of the signal mode is fixed to \SI{11.4}{\THz}, by increasing the pump power $P_p$, we can switch the topological interface mode (originally located inside the bandgap) to a frequency region occupied by leaky bulk modes (outside the bandgap), namely a frequency at which the signal can no longer propagate in the waveguide. The threshold pump power needed to switch off the signal propagation was calculated to be $P_p=\SI{3.25}{\uW}$. It should be noted that we observed that the relative location of the bands did not change significantly as the pump power increased, although, as we just explained, their frequency varied.

\begin{figure}[t!]
\centering
\includegraphics[width=0.4\textwidth]{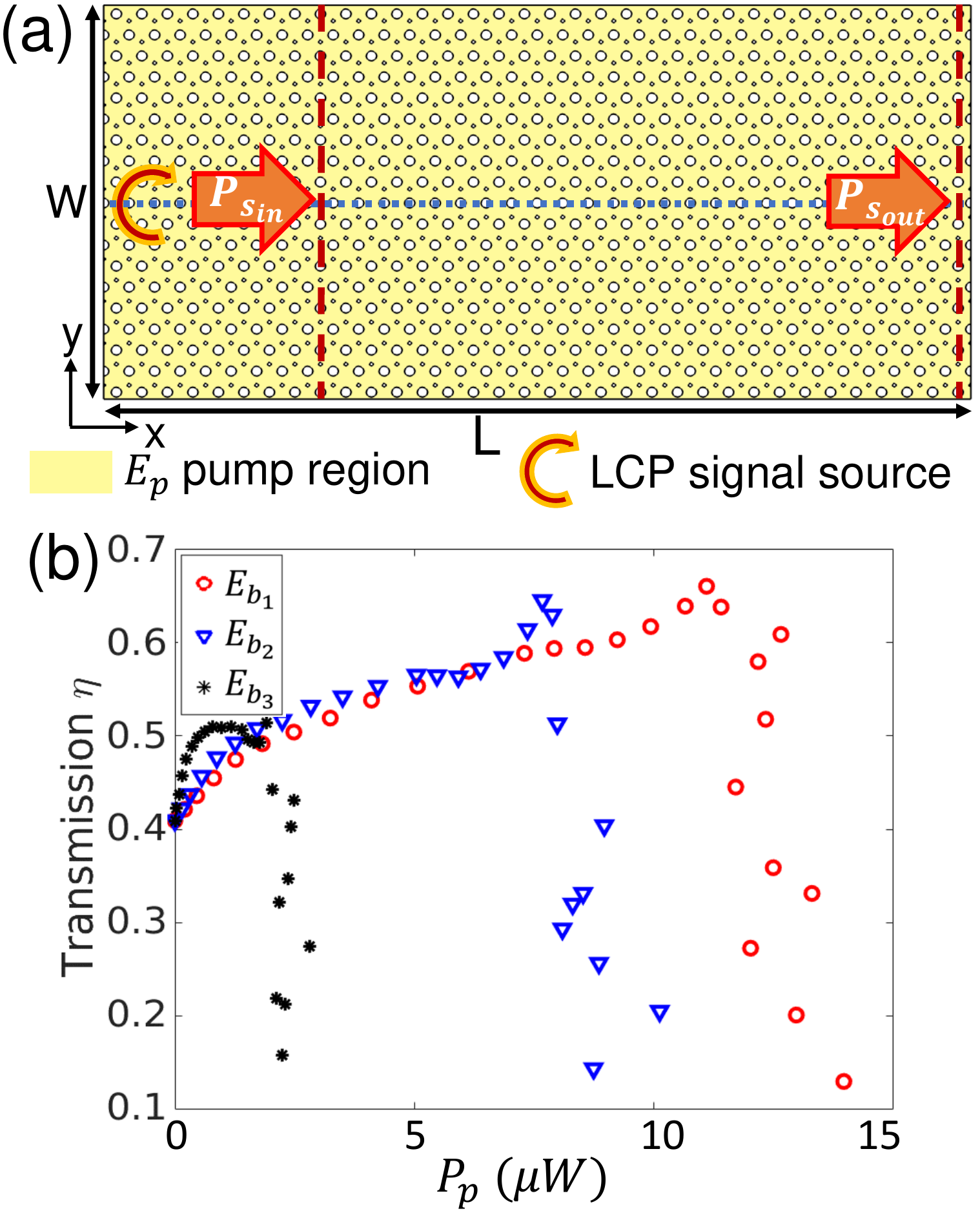}
\caption{(a) Schematic of the active all-optical switch based on the graphene plasmonic waveguide. Light is pumped in the bulk mode $E_b$ (yellow region), whereas the signal $E_{s}$ is generated by a LCP source at $\SI{11.4}{\THz}$ and carries an input power $P_{s_{in}}$ and output power $P_{s_{out}}$. The length of the graphene waveguide is $L=35a$ and the device width is $W=19a$. (b) Signal transmission, defined as $\eta=P_{s_{out}}/P_{s_{in}}$, vs. the pump power $P_p$, determined for the bulk modes with different GV, $E_{b_1}$, $E_{b_2}$, and $E_{b_3}$, indicated in Fig.~\ref{cap3}(a).}
\label{cap5}
\end{figure}

After the validation of the main idea on which our proposed optically controllable switch operates, namely the optical tuning of the band structure of the graphene metasurface via a pump beam, we proceeded to analyze quantitatively the optical characteristics of our optical switch. To this end, we considered the optical device depicted schematically in Fig.~\ref{cap5}(a), which is operated as follows: a pump beam $E_p$ (yellow region) with a frequency chosen in such a way that it propagates in a bulk mode, $E_b$, is injected into the graphene metasurface, while a low power LCP source with frequency $\SI{11.4}{\THz}$ generates an optical signal (probe) $E_s$. The length of the graphene plasmonic metasurface was $L=35a$, which ensured that one achieves a relatively stable signal propagation at the domain-wall interface along the $x$-axis well beyond the transient region where the light is coupled from the LCP source to the signal mode. Due to the fact that the optical signal with a frequency of $\SI{11.4}{\THz}$ can be switched from a topological interface mode to leaky bulk modes by tuning the pump power, the optical signal can be turned ON/OFF when the pump power increases beyond a certain threshold. To validate and quantify these ideas, we define the transmission of the signal beam as the ratio between the output power of the signal, $P_{s_{out}}$, measured at the end of the graphene waveguide, and the input power of the signal, $P_{s_{in}}$, measured nearby the LCP source, namely $\eta=P_{s_{out}}/P_{s_{in}}$. Given that the power of the signal depends on the pump power, the transmission $\eta$ depends implicitly on $P_{p}$, too.

The results of our computational analysis of the proposed optical switch are summarized in Fig.~\ref{cap5}(b), where we plot the dependence of the transmission of the signal power $\eta$ on the pump power $P_p$, determined for the cases when the pump power is injected in the bulk modes with different GV, $E_{b_1}$ (red circles), $E_{b_2}$ (blue triangles), and $E_{b_3}$ (black stars), indicated in Fig.~\ref{cap3}(a). By employing the three pump modes, $E_{b_1}$, $E_{b_2}$, and $E_{b_3}$, with decreasing GV approaching the SL regime, allowed us the investigate the influence of SL effects on the device characteristics. One conclusion of this analysis is that the signal transmission corresponding to pump modes with different GV presents a similar trend. Specifically, the signal transmission increases with the increase of the pump power, which is explained by the fact that, due to the Kerr-effect-induced increment of the refractive index of graphene waveguide, the signal can be more readily focused and coupled from the excitation source into the topological mode. Subsequently, as the pump power is further increased, the signal mode begins to couple to bulk modes resulting in a steep decrease to less than $0.1$ of the signal transmission.

Another important fact revealed by the plots presented in Fig.~\ref{cap5}(b) is that the pump power needed to switch ON/OFF the signal decreases dramatically when the GV of the pump mode is pushed into the SL regime. In particular, more than $6\times$ reduction of the switching optical pump power is observed in our numerical investigations when the pump mode is tuned from $E_{b_1}$ to $E_{b_3}$. This important result can be easily understood from the information conveyed by Eq.~\eqref{eq:2}. Thus, it can be seen from this equation that for a given pump power, $P_{p}$, the smaller the GV, $v_{g}$, of the pump mode is, the larger is the corresponding electric field intensity $\vert \mathbf{E}_{b}\vert$ of the pump mode. This in turn implies that a larger Kerr-induce variation of the index of refraction of graphene is achieved and consequently a larger frequency shift of the photonic bands is attained.

\section{Conclusion}
We demonstrated an active optically-tunable switch based on the Kerr effect in a topologically protected valley-Hall graphene plasmonic metasurface. The spatial-inversion symmetry breaking of the graphene nanohole metasurface is realized by introducing an extra hole, which opens up a symmetry-protected Dirac cone. Consequently, a topological interface mode is formed inside the nontrivial bandgap and propagates along the domain-wall interface in a unidirectional manner. Taking advantage of a large nonlinear refractive index of the graphene, the Kerr effect induced in the graphene is used to shift the frequency of the nontrivial bandgap of the system, so that one can optically control the signal propagation in the proposed topological graphene waveguide. Importantly, our results show that the transmission of the optical signal sharply decreases as the the pump power increases, and the required switching pump power is significantly reduced when the all-optical switch operates in the slow-light regime.

\textbf{Funding}
European Research Council (ERC- 2014-CoG-648328); China Scholarship Council.

\textbf{Disclosures}
The authors declare no conflicts of interest.

\end{document}